# Deep Levels and Mixed Conductivity in Organometallic Halide Perovskites

Artem Musiienko[1*], Pavel Moravec[1], Roman Grill[1], Petr Praus[1], Igor Vasylchenko[1], Jakub Pekarek[1], Jeremy Tisdale[2], Katarina Ridzonova[1], Eduard Belas[1], Lucie Abelová[3,4], Bin Hu[2], Eric Lukosi[5], Mahshid Ahmadi[2*]

[1]*Institute of Physics, Charles University, Prague, 121 16, Czech Republic*
[2]*Joint Institute for Advanced Materials, Department of Materials Science and Engineering, University of Tennessee, Knoxville, TN 37996, USA*
[3] *Institute of Physics, Academy of Sciences of the Czech Republic, Prague, Czech Republic.*
[4] *Electrical Engineering, Czech Technical University in Prague, Prague, Czech Republic*
[5]*Joint Institute for Advanced Materials, Department of Nuclear Engineering, University of Tennessee, Knoxville, TN 37996, USA*

*Corresponding author e-mails:
M. Ahmadi: mahmadi3@utk.edu, A. Musiienko: musienko.art@gmail.com

**Abstract:**

Understanding the type, formation energy and capture cross section of defects is one of the challenges in the field of organometallic halide perovskite (OMHP) devices. Currently, such understanding is limited, restricting the power conversion efficiencies of OMHPs solar cells from reaching their Shockley–Queisser limit. In more matured semiconductors like Si, the knowledge of defects was one of the major factor in successful technological implementation. This knowledge and its control can make a paradigm in development of OMHP devices. Here, we report on deep level (DL) defects and their effect on free charge transport properties of single crystalline methylammonium lead bromide perovskite (MAPbBr$_3$). In order to determine DL activation energy and capture cross section we used photo-Hall effect spectroscopy (PHES) with enhanced illumination in both steady-state and dynamic regimes. This method has shown to be convenient due to the direct DL visualization by sub-bandgap photo-excitation of trapped carriers. DLs with activation energies of $E_V$ + 1.05 eV, $E_V$ + 1.5 eV, and $E_V$ + 1.9 eV (or $E_C$ - 1.9 eV) were detected. The hole capture cross section of $\sigma_h = 4 \times 10^{-17}$ cm$^2$ is found using photoconductivity relaxation after sub-bandgap photo-excitation. Here, we found the DL defects responsible for non-radiative recombination and its impact on band alignment for the first time. Additionally, the transport properties of single crystal MAPbBr$_3$ is measured by Time of Flight (ToF) at several biases. The analysis of ToF measurement further confirms increase of Hall mobility and the enhancement of hole transport produced by sub-bandgap illumination in MAPbBr$_3$ devices. Here, our studies provide a strong evidence on deep levels in OMHPs and opens a richer picture of the role and properties of deep levels in MAPbBr$_3$ single crystals as a system model for the first time. The deeper knowledge of the electrical structure of OMHP could open further opportunities in the development of more feasible technology. Indeed, knowledge of the exact position of DLs is beneficial in controlling these defects by crystal growth modification to eliminate these defects as was done for classical inorganic semiconductors including in Si, GaAs and CdTe development.

**Introduction**

Since the important work of Kojima *et al.*[1] in 2009 showing the power conversion efficiency (PCE) of around 3.81% for solid-state perovskite solar cells, organometallic halide perovskites (OMHPs) have experienced a remarkable development as highly efficient optoelectronic materials for a variety of applications particularly solar cells[2], light emitting diodes[3] (LEDs) and photodetectors[4]. Recently, it has been shown by several groups that the outstanding intrinsic properties in OMHPs single crystals can empower a new generation of ionizing radiation sensors with remarkably high mobility-lifetime product[5,6]. One notable property of OMHPs is the low-temperature solution growth, which is considerably lower than high-temperature melt growth techniques commonly used for classical inorganic semiconductors (e.g., Si, CdTe, GaAs, Ge, etc.). However, one can expect a relatively high concentration of defects introduced during low-temperature growth of single crystal OMHP and its polycrystalline thin films. So far, there is limited information on the properties and electronic impact of defects in OMHP device performance. In general, defects can change the electric field distribution and charge transport in a device and/or response time, and reduce device signal integrity. The semiconducting properties, controlled by shallow and deep traps, can be described from three parameters: i) majority charge carriers, ii) mobility, and iii) charge carrier lifetime. Based on impedance spectroscopy and numerical analysis, it has been shown that OMHPs are, in fact, intrinsic semiconductors, robust against accidental extrinsic doping[7]. This is attributed to the physicochemical mechanism related to band structure, requiring a large formation energy of deep traps within the band structure, effectively protecting the perovskite semiconductor from being extrinsically doped. While shallow level defects can be ionized at room temperature and act as a dopant which effects free carrier concentration and conductivity, deep level defects can pin the Fermi level and limit charge carrier transport. In more matured semiconductors like Si, the knowledge of defects was one of the major factors in successful technological implementation[8]. Therefore, understanding the type, formation energy, and capture cross-section of defects is critical and can lead to further development of OMHP devices.

Several studies explain that main intrinsic defects are shallow level defects leading to long carrier diffusion length and carrier lifetime coupled to efficient OMHP devices suggesting a class of materials with low concentration of defects and/or highly defect tolerance[9–11]. The low concentration of deep level defect is attributed to Schottky type defects created when a quasi-neutral space charge (SC) results from a near-equal concertation of anion and cation vacancies in the material. However, it was revealed that the performance of OMHP devices is mainly limited by Shockley-Read-Hall (SHR) recombination confirming that defects are dynamic elements and have a critical role in the development of OMHP devices[9,12,13]. Although maybe benign, it is crucial to fundamentally understand how deep levels can explain the long carrier lifetime and how this characteristic can be used in the practical development of OMHP devices.

Non-radiative, defect-assisted carrier recombination is particularly important for operation of solar cells with higher $V_{oc}$ and in achieving high charge collection efficiency in ionizing radiation sensors[14]. According to DFT calculations, deep level defects in OMHPs have high formation energies and are related to the growth condition[15]. It was predicted that Pb and halide interstitials and antisites can create deep level defects [9,12,15–17]. The shallow acceptor defects with low formation energies were identified as $V_{Pb}^{-2}$, $V_{MA}^{-1}$, $I_i^{-1}$, and the shallow donor defects as $V_I^{+1}$, $MA_i^{+1}$, $Pb_i^{+2}$ [9,18,19] which can intrinsically dope OMHPs as p-type or n-type, respectively[9,16]. Still, experimental verification of computationally predicted deep level (DL) states within the bandgap are needed to further improve OMHP device performance.

Several different deep level spectroscopy methods have been used to identify deep-level ionization energy[20–24] and capture cross section[20,25] in conventional inorganic semiconductors. These methods face serious obstacles in organic and hybrid semiconductor materials due to low DL concentrations and capture cross section[18,26], and therefore the electrical structure of such organic semiconductors remains poorly understood. It is also a common problem in both organic and hybrid semiconductors to identify relative deep level (DL) positions inside the band gap, i.e. whether DL energy locates below or above the Fermi energy. The knowledge of DL positions inside the band gap plays a crucial role in semiconductor development[25] and therefore, appropriate methods must be applied to identify DLs in OMHPs. Chen *et al*. have shown that Hall effect and photoconductivity measurements are an effective technique in studying charge transport and carrier trapping in OMHP devices[27] and organic field-effect transistors[28]. In this method, Chen *et al*. applied a photoexcitation with photon energies higher than band gap to generate free electrons and holes. The generation of both carrier types lead to a low Hall signal. Therefore, they used a rotating magnetic field in combination with lock-in amplifier to amplify the signal. However, this can complicate the Hall measurement set up. In addition, light illumination with photon energies higher than band gap energy can mix holes and electrons signals which makes information on a single carrier property unavailable. Here, we develop photo-Hall effect spectroscopy (PHES) technique without rotating magnetic field and based on sub-bandgap photoexcitation to enhance Hall signal due to single carrier generation from DLs. The variation of photon energy can further be used to probe the electrical structure of the material. Recently, we have shown that PHES modified by amplified sub-bandgap illumination can be used as an effective tool to study DL defects, including the properties of recombination-like levels in CdTe based detectors[29,30].

In general, PHES is a straightforward technique based on the direct visualization of deep states by the free carriers' excitation. In this method, the enhanced photon flux produced by a filtered white laser is used to generate free charge carriers via sub-bandgap photo-excitation of DLs where low absorption is typically observed[26,31]. The PHES measurements can be divided into three main groups: i) the Hall signal measurements with different illuminating photon energies which allows DL threshold energy detection, ii) the Hall mobility and photo-Hall conductivity (PhC) as a function of illumination intensity, and iii) the relaxation time of photoconductivity allowing determination of DL recombination properties. In this manuscript we examine the electrical structure, mixed conductivity, and recombination properties of $MAPbBr_3$ single crystals by PHES in both steady-state and dynamic regimes to determine the energy and capture cross section of DLs in this material for the first time.

Further we apply current waveforms (CWF) time of flight (ToF) measurements to investigate the transport properties of $MAPbBr_3$ single crystals. Time of flight (ToF) technique has been widely used to study charge carrier drift mobility in OMHP devices[32–36]. The CWF recorded in ToF are induced by photo-generated charge carriers drifting through the sample under applied biases. When the above-bandgap laser light pulses are used for charge carriers generation several transport parameters of semiconductors can be evaluated including transit time, charge carrier mobility and lifetime of free carriers. The drift mobility measured by ToF is used to examine mixed conductivity in $MAPbBr_3$ single crystal device.

**Results**

**Deep-level parameters detected by photo-Hall effect spectroscopy (PHES)**

A scheme of the PHES measurements and an image of the device tested are provided in **Figure 1(a)** and **Figure 1(c)**. The DL photo-excitation mechanisms are presented in **Figure 1(b)**. The monochromatic illumination activates defects creating free charge carriers. The type of charge carrier is dependent on the nature of the activated deep level. The photo-generated carriers are affected by Lorentz forces, and the observed changes in Hall measurements enable the evaluation of the DL trap properties. The bipolar Hall mobility is given by the relation

$$\mu_H = \left(A_p\mu_p^2 p - A_n\mu_n^2 n\right) \Big/ \left(\mu_p p + \mu_n n\right) \qquad (1)$$

where $\mu_e$, $\mu_p$, $n$, and $p$ are electron mobility, hole mobility, electron, and hole concentrations, respectively. The scattering parameters $A_p$ and $A_n$ are constants, and here we use the assumption that $A_p = A_n = 1$. According to Eq. 1, the energy level of the DL traps inside the band gap can be detected by the observed change in the Hall mobility and photoconductivity. However, due to higher quality of PhC signal compare to Hall mobility signal, which is few orders of magnitude lower, the detection of DL activation energy is usually based on photoconductivity spectra. When a photon energy is enough to activate deep levels, free carriers, electrons or holes, are generated. The lowest energy where PhC response or enhancement is detected considered as activation energy. Note that PhC does not contain information on the nature of photo-exited free charge carriers and as a result on the relative position of DLs in the band gap. Therefore, the Hall mobility in PHES is used to show free carrier nature, whether free holes or electrons are generated. The Hall mobility increases in a *p*-type material due to free hole generation assisted by photo-excited acceptor-like DL, model 2, and the free electron generation assisted by photo-excited donor-like DL, model 1, leads to a decrease in the Hall mobility; both processes are shown in **Figure 1(b)**. Note that free electron generation also leads to a conversion of the Hall voltage from positive to negative sign.

As it was mentioned earlier, the detection of deep level states in OMHPs is a known challenge. In order to discover the electrical structure and to find the main DL properties, we studied single crystal MAPbBr$_3$ devices by photo-Hall effect spectroscopy with enhanced sub-bandgap illumination. The intensity of light used was 1 mWcm$^{-2}$, significantly less than that used in photovoltaic operation[37], and does not lead to a photo-induced material degradation[38]. Single crystals typical dimensions are around 1×2.5×9 mm$^3$. Details of crystal growth can be found in materials section. The photoconductivity and Hall mobility spectra in the energy range of 1.0-2.4 eV are shown in **Figure 2**. Two single crystal samples show very similar PHES results. The MAPbBr$_3$ crystals show no photocurrent response upon illumination with a photon energy less than 1.05 eV, which is where the first DL threshold energy is observed. The second and third threshold energies are detected at 1.5 eV and 1.9 eV, respectively. Note that PhC in **Figure 2 (a)** is shown in logarithmic scale to present the whole spectrum. The PhC spectra in a linear scale can be found in supplementary information, **Figure S1**. As can be seen in **Figure S1** the initial photon energies started from 0.6 eV when the first increase is detected at 1.05 eV.

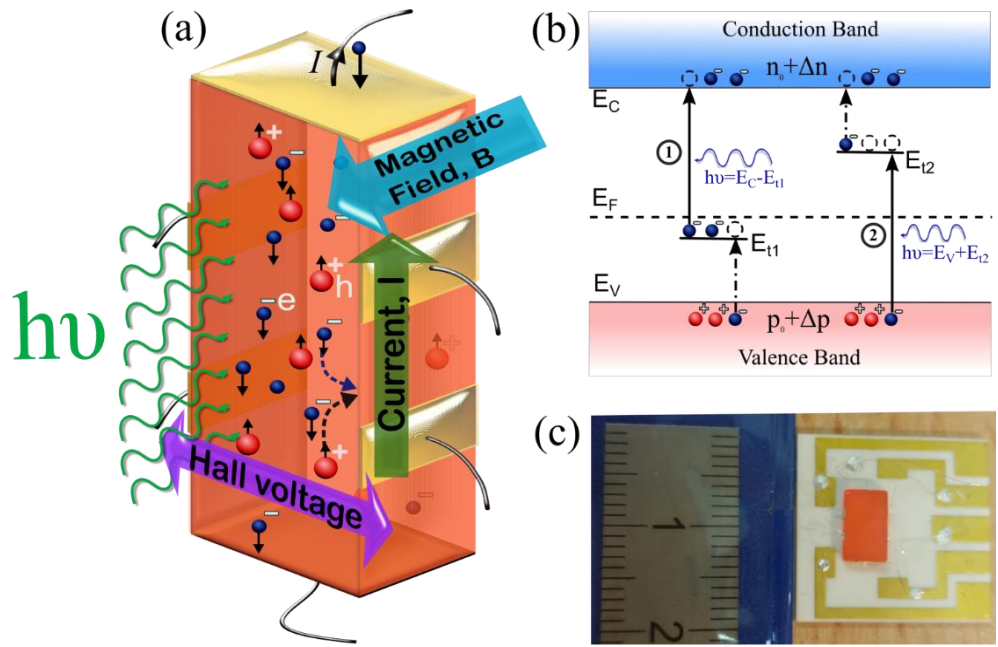

**Figure 1** (a) The basic physical principles of PHES method. (b) Competing deep level models 1 and 2 in the case of the donor-like and acceptor-like DLs, respectively. Solid and dashed arrows show direct and secondary carrier generation processes correspondingly. $E_F$ shows Fermi level position in *p*-type material. (c) A photo of typical solution-grown MAPbBr$_3$ single crystal with Au contacts in a 6-probe/Hall bar geometry. The scale shows centimeters.

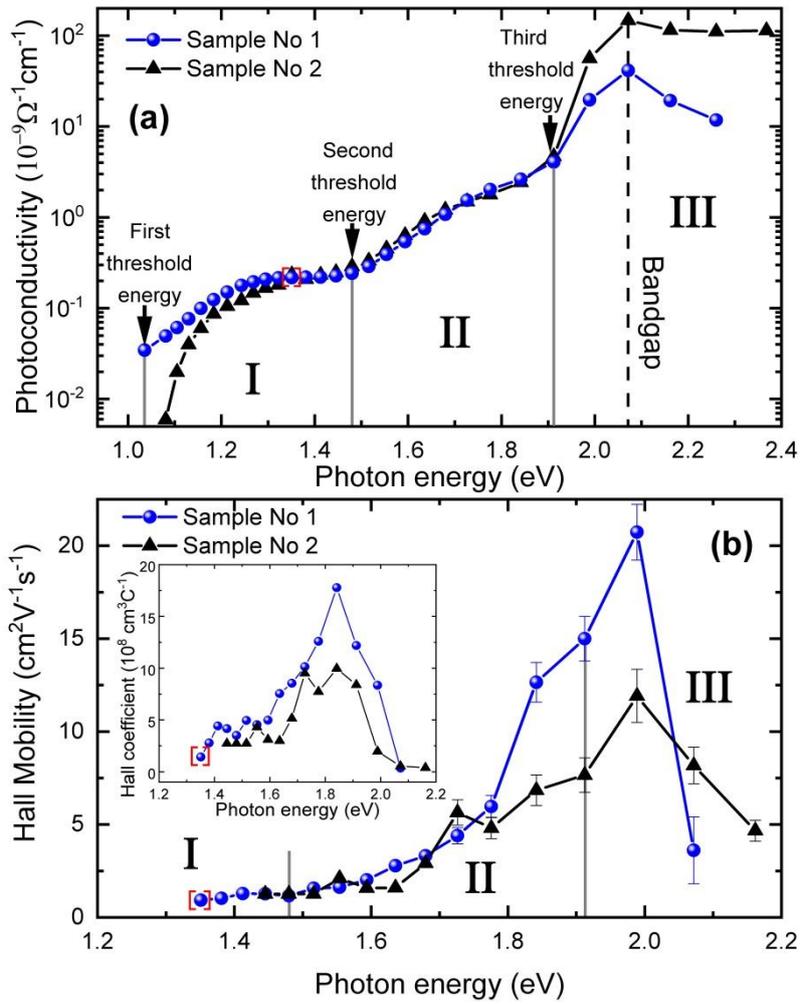

**Figure 2** (a) Photoconductivity and (b) Hall mobility spectra of two single crystals of MAPbBr$_3$ as a function of photon energy. Vertical arrows show the deep level threshold energy. The inset in Figure 2 (b) shows the corresponding Hall coefficient as a function of the photon energy.

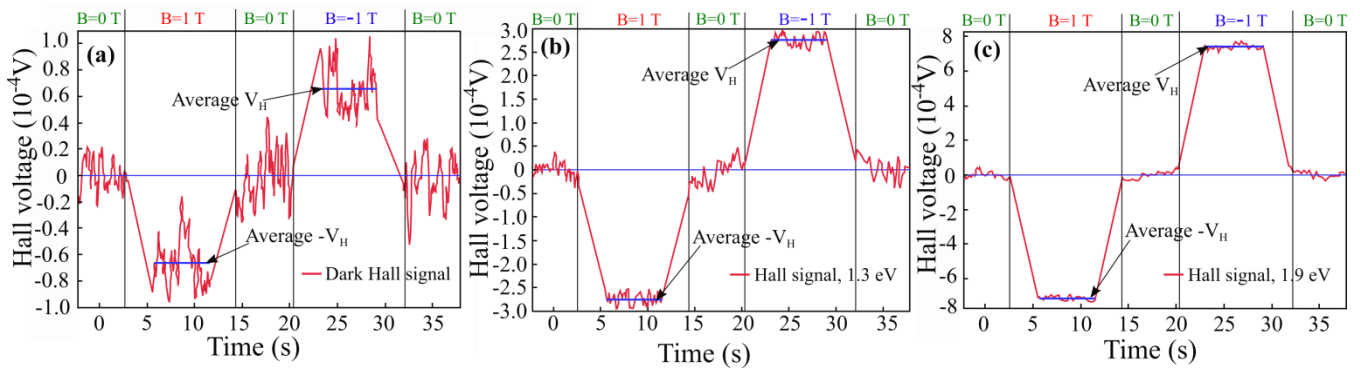

**Figure 3** (a) The evolution of the Hall voltage measured (a) in dark condition, (b) under illumination with a photon energy of 1.3 eV, and (c) under illumination with a photon energy of 1.9 eV for sample No 1. A photon flux of $10^{15}$ cm$^{-2}$s$^{-1}$ is used for both photon energies. Vertical grids represent time regions where the magnetic field is set to 0 T, 1 T, and -1 T, respectively.

Here, MAPbBr$_3$ sample shows a poor Hall signal in dark condition as illustrated in **Figure 3(a)**, where a signal to noise ratio of 0.7 is observed. The Hall voltage signal is significantly enhanced by sub-bandgap illumination as evident from **Figure 3(b)-(c)**. The Hall mobility and Hall coefficient spectra are presented in **Figure 2(b)**. The reliable Hall effect data can be extracted starting from the photon energy of 1.35 eV where the Hall mobility of 1.4 cm$^2$V$^{-1}$s$^{-1}$ is measured. The Hall mobility enhancement correlates with an increase in photoconductivity, and the second DL threshold energy coincides with threshold energy detected from the photoconductivity spectrum. The Hall coefficient $R_H$ also increased under sub-bandgap illumination as shown in the inset in **Figure 2(b)**. The Hall voltage, and as a result the Hall coefficient, shows the positive sign of the polarity which is evidence of the dominant hole concentration and *p*-type conductivity of the material. The photoconductivity and $\mu_H$ enhancement is therefore assigned to free hole generation. Note that increasing of the Hall coefficient contradicts the classical single carrier relation $R_H = 1/(q_e p)$ where the increase of the free hole concentration, *p*, leads to a decrease in $R_H$. The alternative model, the parallel free electron activity[27], also fails to explain the Hall mobility and Hall coefficient increase, especially in sub-bandgap illumination condition where only one type of carrier is generated. Such deviation can be explained by mixed ionic-electronic conductivity in this class of materials and needs to be carefully considered.

The Hall coefficient and conductivity can be explicitly found from the solution of the Boltzmann[39] equation where all moving carriers, including low mobile ions, are considered in the model. The mixed conductivity, *σ*, of a *p*-type material with dominant ion concentration, $N_i$, can be presented by the following equation

$$\sigma = q_e \left( \mu_p p + \mu_i N_i \right) \tag{2}$$

where $\mu_i$ and $q_e$, are ion mobility, and elementary charge respectively. The Hall mobility $\mu_H$, and the Hall coefficient $R_H$ can also be modified considering ionic conductivity as follows:

$$R_H = \frac{1}{q_e p \left(1 + \frac{\mu_i N_i}{\mu_p p}\right)^2} \tag{3}$$

$$\mu_H = \mu_p \frac{1}{1 + \frac{\mu_i N_i}{\mu_p p}} \tag{4}$$

In the case of a material with multiple types of ions, we assume that a single ion type has a maximal $\mu_i N_i$ product which dominates in Eqs. (2-4). If the hole drift mobility $\mu_p$ is known, Eqs. (2-3) can be further resolved to find the hole carrier concentration (*p*) as well as $\mu_i N_i$ product. In this study, we do not differentiate ions from mobile defects[19] (or defect + polaron system[40]) which may both be present in OMHPs, and the term "ion" is applied in the sense of low mobility charge carriers.

As evident from Eq. (2), the increase of the hole concentration, $p = p_0 + \Delta p$, created by light induced photo carriers $\Delta p$, leads to the increase of $R_H$ in a material with mixed conductivity as a result of much higher hole mobility. As the value of $R_H$ is proportional to $V_H$ the free hole generation leads to the Hall signal enhancement as is also theoretically predicted by Eq. 3 and experimentally proven in **Figure 3(b)-(c)**. The detailed consideration of the Hall coefficient and mobility enhancement is shown in supplementary information, **Figure S2** and **S3**. The lower energy Hall data in **Figure 2(b)** are affected by the noise and ions suppression due to the low concentration of free holes which can produce the Hall signal. According to Eq. (4), under intense illumination, the value of $\mu_H$ overwhelms ionic contribution due to photo-induced free hole generation and strive to the value of $\mu_p$. Note that PHES deep level detection concept based on the direct visualization of deep states by the free carriers' excitation is still valid for a material with mixed electronic/ionic conductivity. Because of this free hole generation, the threshold energies detected from photoconductivity and Hall mobility spectra can be assigned as deep states D1 and D2 with activation energies $E_V + 1.05$ eV and $E_V + 1.48$ eV, as shown in **Figure 4**. The energy regions I and II indicated in **Figure 2** correspond to the sub-bandgap light excitation of DLs, D1 and D2, respectively. The increase in photoconductivity in these energy regions is attributed to an increase in available density of states (roughly DOS~$\sqrt{hv}$ ) in the edge of the valance band[41]. The photoconductivity saturation tendency at 1.3-1.43 eV can be explained by a known dependence of a photon capture cross-section on photon energy[42,43] and also by the depletion of exited DL. The notable rise in the photoconductivity in energy region III ($hv > 1.9$ eV) are normally recognized as shallower level[38], D3, and phonon-assisted band-to-band absorption. The interpretation of DL D3 is more complicated due to the proximity of the level to one of the bands. The excitation of D3 and consequent hole generation can be achieved by two excitation mechanisms. The first one, direct photon absorption and electron transition to DL with activation energy $E_V + 1.9$ eV followed by free hole generation. The second mechanism is based on the photo-excitation of DL D3 (alternative model) with activation energy $E_C - 1.9$ eV where hole generation is reached by thermal and secondary photo-excitation of electrons in the valence band as shown in **Figure 4**. The shallow states with activation energies of $E_t < 0.3$ eV were also detected in previous studies[44,45]. The drop in photoconductivity for the incident light above 2.1 eV is associated with bandgap edge absorption where the light is preferably absorbed on the surface. The bandgap energy of 2.2 eV is estimated considering the spectral resolution of the light source in that energy region, which is in agreement with UV-Vis and photoluminescence emission spectra shown in **Figure S4** in the methods part of the paper.

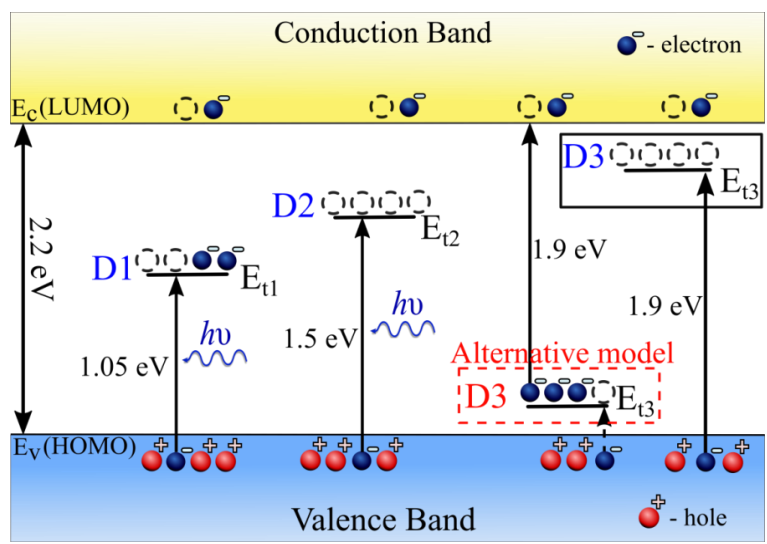

**Figure 4** Experimentally estimated DL transitions in MAPbBr$_3$. Full upward black arrows delineate principal optical excitation. The dashed arrow shows the alternative secondary hole generation process.

The samples were originally grown in the lab in the University of Tennessee, Knoxville, USA. To ensure that the visualized DLs do not belong to these specific samples, the authors grow MAPbBr$_3$ single crystals with slightly different growth method in the lab of Czech Technical University in Prague (details can be found in materials section). Despite the lower quality of the single crystals in the second method and as result a lower resistivity, the measured spectra show the same trends and same DL activation energies (Supplementary Information, **Figure S5**) confirming the inherent properties of MAPbBr$_3$ single crystals.

**Mixed ionic and electronic conductivity and transport properties of MAPbBr$_3$**

Due to the same trends in PHES results for the all samples, hereinafter sample No 1 is chosen for more detailed study. To evaluate the ionic contribution to the measured conductivity under high illumination, photoconductivity and Hall mobility (Hall coefficient) were studied as a function of the photon flux at 1.46 and 1.9 eV photon energies for sample 1 (see **Figure 5(a-b)**). Both the photoconductivity and $\mu_H$ show a super-linear rise in response to illumination intensity. The maximal value of $\mu_H$, 87.1±2.5 cm$^2$V$^{-1}$s$^{-1}$, is attained at 1.92 eV photon energy and a photon flux of $1.7\times10^{-16}$ cm$^{-2}$s$^{-1}$. The relationship between the photoconductivity on the photon flux $I$ can be estimated by ~$I^\alpha$ with $\alpha \approx 1.5$. The super-linear rise of $\mu_H$ indicates that actual hole mobility can be higher than maximal value reached by PHES set-up. To find the hole mobility, time of flight (ToF) method was applied.

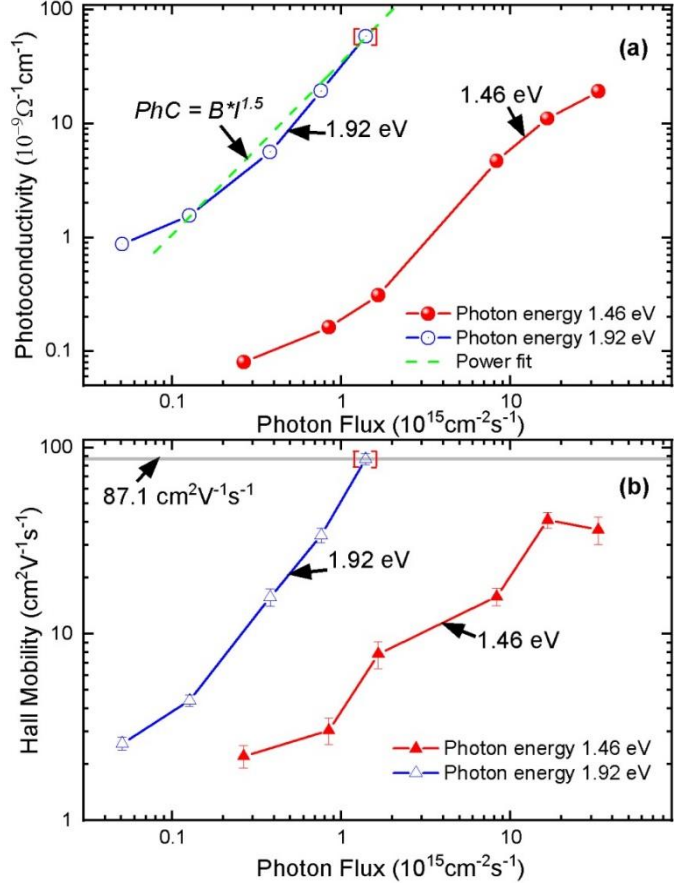

**Figure 5** (a) Photoconductivity and (b) Hall Mobility as a function of photon flux at 1.46 eV and 1.92 eV high-flux illumination in MAPbBr$_3$. The horizontal line in Figure 5(b) represents the maximum Hall mobility.

The semiconductor characterization by current waveforms (CWF) ToF is based on the carrier generation by a short above-bandgap laser pulse focused at the center of the anode. The photo-generated free holes are collected on the negative biased cathode. The current transient induced by moving holes is expressed by a single exponent relation[46]

$$i(t) \propto e^{(-t/\tau)}, \tag{5}$$

where $\tau$ is holes lifetime characterizing their passing through the device volume. As can be seen in **Figure 6**, a constant profile of the electric field is found in the device which can be further verified from the transit time electric filed dependence. The hole mobility $\mu_p$ can be found from the transit time $t_r$ and the bias $U$ according to

$$\mu_p = \frac{L^2}{t_r U} \tag{6}$$

where *L* is a detector thickness. The transit time is found from the inflection point of the current waveform which gives a typical error of 10 % in the mobility value[47].

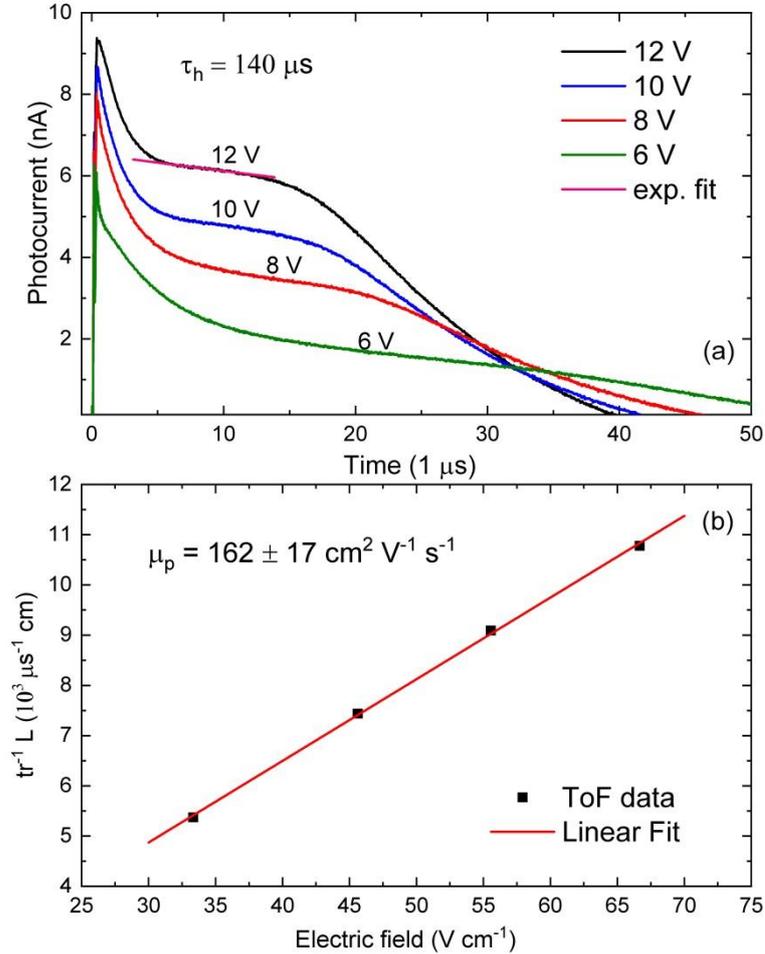

**Figure 6 (a)** CWF ToF measurements of MAPbBr$_3$ single crystals with different applied voltage. **(b)** Transit time as a function of applied voltage.

The transit time represents the time needed for photo-induced holes to pass through the whole device, and typically $t_r \neq \tau$. Here, the calculated ToF hole mobility is 162±17 cm$^2$V$^{-1}$s$^{-1}$ and the hole lifetime deduced from the transit region and Eq. (5) is 140 µs. The value of lifetime measured by ToF is mainly controlled by DLs. Here, the bimolecular recombination is low due to only hole transport and low electron concentration.

Since the Hall data can be extracted using light illumination, the Hall relations in Eq. (2-3), which include ionic contributions, may be used to find $p$ and $\mu_i N_i$. It is known that for semiconductors and for MAPbBr$_3$, in particular, the drift mobility is affected by phonons or polarones[48–51]. To activate additional scattering mechanisms (charged and neutral defects) either a large concentration of charged defects should be present, or a high photoexcitation intensity should be applied. Therefore, the limitation of drift mobility caused by the enhanced ionized defect scattering is not significant at room temperature when the ionized defect density is below $10^{17}$cm$^3$ [52,53]. Consequently, the constant drift mobility can be safely used in the further calculations. The value of the hole mobility found by ToF is used in the calculations. The data in **Figure 5** were used to find ionic and free carriers contributions at high illumination intensity, and the data in **Figure 2** were used in the case of low illumination rate. The corresponding points which were used for calculation are highlighted with red brackets in **Figure 2**. The ionic parameter $\mu_i N_i = 2\times10^{11}$ cm$^{-1}$V$^{-1}$s$^{-1}$ and the hole concentration $p = 8\times10^8$ cm$^{-3}$ is found at the point of maximal Hall mobility. The ionic impact in the overall conductivity is estimated to be around 62 % under high illumination intensity and increases up to 98 % at a low light intensity where $\mu_i N_i = 2\times10^{10}$ cm$^{-1}$V$^{-1}$s$^{-1}$ and $p = 6\times10^5$ cm$^{-3}$. We assume that the value of $\mu_i N_i$ and $p$ at low illumination intensity slightly deviates from their actual values in the dark. The ion mobility is estimated to be in the range of $10^{-10}$-$10^{-7}$ cm$^2$V$^{-1}$s$^{-1}$ considering ions[54] or mobile vacancies concentration in the range $10^{17}$-$10^{20}$ cm$^{-3}$ calculated by Frost et al.[55]. The diffusion coefficient of $10^{-12}$-$10^{-9}$ cm$^2$s$^{-1}$ is estimated from the found ionic mobility. The contribution of ions in Hall measurements has been explained in detail in Supplementary information.

**Deep level recombination parameters**

The Shockley-Reed-Hall (SRH) theory and photoconductivity relaxation can be used to determine the capture cross section of DL traps[29,30]. The kinetic equation of SRH hole recombination rate for a single DL can be found using Eq. (7), which describes hole relaxation after continuous excitation is switched off.

$$\frac{\partial p}{\partial t} = -\sigma_h v_h p(N_t - p_t) \qquad (7)$$

In Eq. (8), $N_t$, $p_t$, and $v_h$ are the total DL concentration, the hole density at the DL, and the thermal velocity of holes, respectively. Assuming holes previously excited from a single DL, which was filled by holes in the dark, the charge neutrality requires that Eq. (8) be fulfilled, where $p_0$ is the hole density in the dark.

$$N_t - p_t = p - p_0 \qquad (8)$$

Substituting Eq. (8) into Eq. (7) we obtain the solution for $p$

$$p = \frac{p_0 p_{max}}{p_{max} + (p_0 - p_{max})e^{-\sigma_h v_h p_0 t}} \qquad (9)$$

where $p_{max}$ is the hole density at the time when photo-induced excitation ceases. The hole density parameters $p_0$ and $p_{max}$ are defined by the experiment using measured conductivity and taking into account the ionic conductivity, 98%, at low illumination intensity identified previously. By using the calculated hole mobility from ToF measurements (162 cm$^2$V$^{-1}$s$^{-1}$), $v_h$ from the well-known formula[56], and the effective hole mass of $0.11 m_e$[57,58] for OMHPs, the hole capture cross section $\sigma_h$ remains the only unknown parameter.

Photoconductivity relaxation measured at 1.13 eV, 1.23 eV, 1.41 eV, and 1.67 eV, where DL excitation is observed, are presented in **Figure 7**. The characteristic time of the hole density relaxation $\tau_h > 20$ s was deduced from the light ON/OFF photoconductivity. The long lifetime of photo-induced holes is assigned to hole recombination with D1 ($E_{t1}$) that has a rather low hole capture cross-section $\sigma_h$. The relaxation of photoconductivity is used to determine $\sigma_h$ of the excited DL. The kinetic equation describing the hole relaxation after the excitation was switched off is expressed as the single parameter fit shown by the red solid line in **Figures 7 (a)-(c)** revealed a $\sigma_h = 4 \times 10^{-17}$ cm$^2$. This is valid for 1.13 eV, 1.23 eV, and 1.41 eV photon energies proving a single DL, denoted as D1, in region I (see **Figure 2**). Note that found capture cross section is obtained from non-thermal measurements. Therefore the obtained value of capture cross section reproduce the real capture of the free carriers[25]. This "real value" of capture cross section is hardly accessible from the other methods because it is typically founded from temperature dependencies[59,60].

More complicated relaxation dynamics is observed at 1.67 eV photon energy excitation and the observed photoconductivity transients cannot be fitted by the single DL model as shown in **Figure 7(d)**. However, part of photoconductivity transient may be fitted by the single DL model, resulting in the same $\sigma_h$ as in **Figure 7(a)-(c)**, indicating that both are electrically active in photoconductive decay observations in energy region II. Therefore, Eq. (9) cannot be used to find hole capture cross section of DLs D2 and D3. On the other hand, positions of these DLs in the bandgap principally limit significant hole recombination or trapping. The activation energy of DL D3 is too close to one of the bands. DL D2 locates far above Fermi level and as a result, it is completely empty.

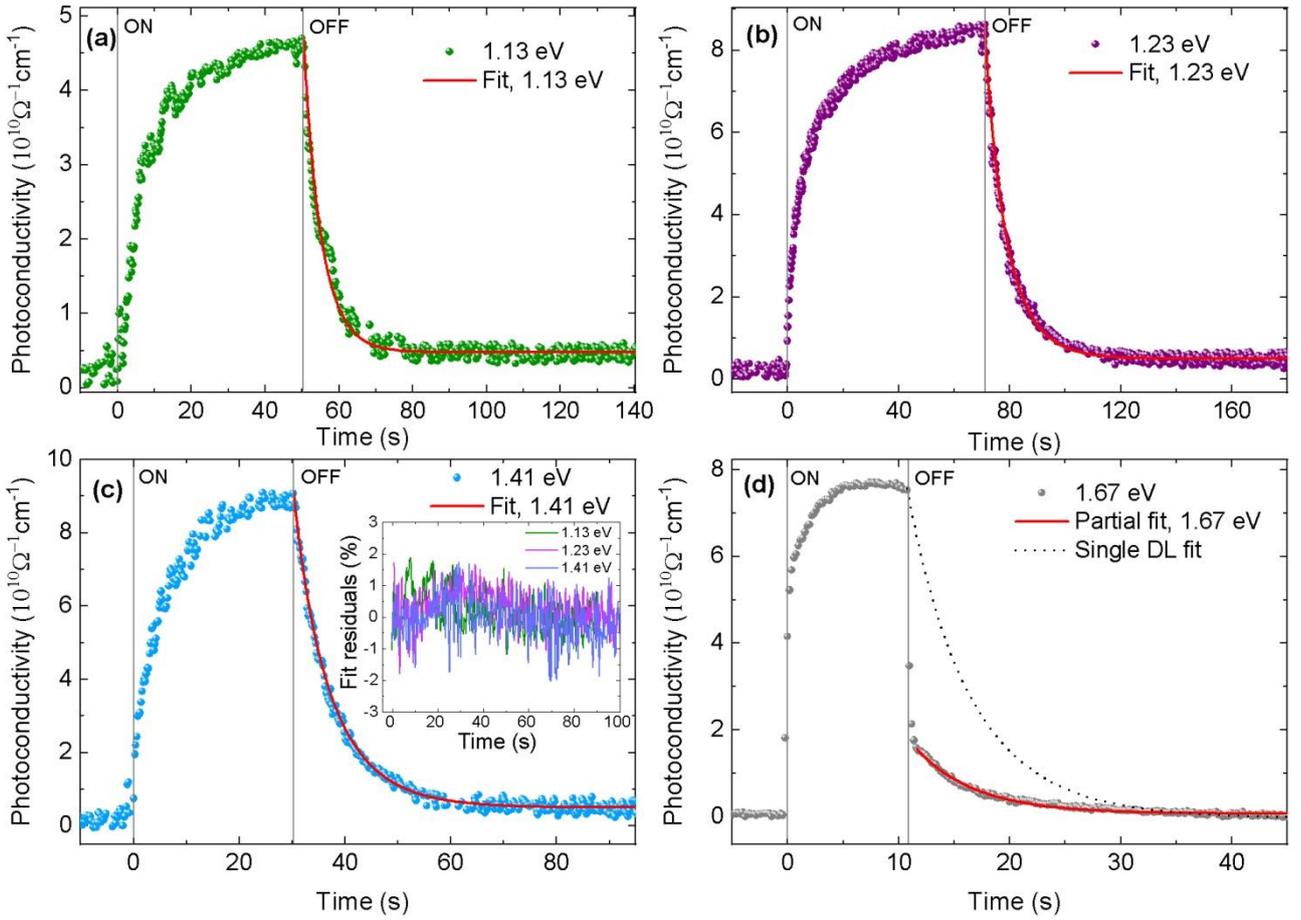

**Figure 7** The photoconductivity relaxation after sub-bandgap illumination at (a) 1.13 eV, (b) 1.23 eV, (c) 1.41 eV, and (d) 1.67 eV photon energies measured in MAPbBr$_3$. Vertical grids represent time regions where the light is turned ON/OFF. The circular symbols represent experimental points and the red line is the fit according to the presented single DL model. The inset in Figure 7 (c) shows the fit residuals.

**Discussion**

In this work, we provide a profound insight into crucial electrical and transport properties of OMHPs based on single crystal MAPbBr$_3$ device. We show that Hall measurements may be enhanced by sub-bandgap illumination producing free holes supplied by the deep levels. The positive sign of Hall voltage and *p*-type conductivity was detected. The maximum value of $\mu_H$, 87.1±2.5 cm$^2$V$^{-1}$s$^{-1}$ was observed via PHES. Due to a super-linear Hall mobility increase, the real value of the hole mobility is expected to be higher and closer to the value found by ToF measurements, 162 cm$^2$V$^{-1}$s$^{-1}$. Theoretical calculations predicted that the carrier mobility limited by acoustic phonon scattering can be several thousand cm$^2$ V$^{-1}$ s$^{-1}$ in OMHPs[61]. The dipole screening[62] or large polaron formation[32,36,57] were proposed to explain such a discrepancy between theoretical and experimental results.

Using the PHES measurements with tunable photon energies, we revealed the intra-bandgap electrical structure of MAPbBr$_3$ single crystals. Deep levels with activation energies $E_V$ + 1.05 eV (D1), $E_V$ + 1.5 eV (D2), and $E_V$ + 1.9 eV (or $E_C$ - 1.9 eV) (D3) were detected. The bandgap of 2.2 eV could be estimated from PHES spectra, which was in agreement with other optical measurements (**Figure S**4). The free hole generation was experimentally proven in **Figure 2(b)** in energy regions I and II; therefore, DLs D1 and D2 energy levels were classified relatively to the valence band. The position of DL D3 with high activation energy remained unclear because of possible interference of optical excitation and excessive thermal excitation of charges from the respective band nearby. The straightforward determination of electrical character of identified DLs is impossible in the single PHES. Both donors and acceptors are found distributed overall in the band gap in semiconductors. Due to the novelty of observed DL transitions we do not cogitate on the microscopic structure of detected DLs in this paper.

The measured photoconductivity in the sub-bandgap energy region is a few orders of magnitude lower than that observed in inorganic semiconductors[29,30,63,64]. The lower photoconductivity may indicate the low defect concentration in the range of $10^9$-$10^{12}$ cm$^{-3}$ according to SRH simulations[30]. The hole capture cross section of $\sigma_h = 4 \times 10^{-17}$ cm$^2$ of DL D1 with the activation energy $E_V$ + 1.05 eV was found from photoconductivity relaxation measurements. The assignment of DLs as electron/hole trap or recombination center is complicated due to unknown electron capture cross section, but few important remarks can be given. It is apparent that, in the presented DL model, hole recombination is controlled by DL D1. Recalling the hole capture cross section and DL concentration, the SRH lifetime of $10^{-4}$-$10^{-1}$s is estimated (according to Eq. 20 in the Supplementary section) which is in good agreement with previous findings[65,66] and the lifetime value 140 μs founded by ToF method in this manuscript. The evaluated free hole lifetime is comparable with the lifetime of free carriers in conventional inorganic semiconductor materials[25,27,67]. Deep levels D2 and D3 are empty in the dark and the electron lifetime can be controlled by one of these deep levels. The photoconductivity super-linearity observed in this study is commonly explained for inorganic materials by multiple DLs in the material[68,69] and cannot be observed by a single DL excitation. The power dependence with $\alpha \approx 1$ was reported by Chen *et al* in MAPbI$_3$[27] and considered as trap assisted recombination with one DL recombination channel involved. In the case of multiple DL recombination channels, the value of $\alpha$ can be greater than 1 and the photon flux dependence can principally show a more complex profile as predicted by SRH theory[30,68–70].

Here, we cannot identify the chemical nature of each DLs. There is controversy over the chemical nature of defects based on several DFT calculations. For example, previous theoretical calculations have estimated that Pb interstitials $Pb_i^{+2}$, Br vacancies ($V_{Br}^+$) and antisite ($Br_{Pb}$) are the dominant deep level defects in MAPbBr$_3$ perovskites[15]. By modifying the calculations using hybrid density functional calculations, Du showed that only halide interstitials ($I_i^-$ in case of MAPbI$_3$) and its complexes (antisite $I_{MA}$ which is not stable and decomposes into $I_i^-$ and $V_{MA}^-$) induce deep electron and hole trapping levels inside the band gap, acting as nonradiative recombination centers[71]. Very recently, Meggiolaro *et al.* have demonstrated that only less abundant interstitial iodine defects are responsible for deep traps in MAPbI$_3$ leaving only short-living hole traps[72]. On the other hands, the possibility of formation of deep hole polarons was also reported which can effect on the charge carrier dynamics[73].

The Mixed ionic-electronic conductivity was demonstrated in MAPbBr$_3$ by PHES measurements. The ionic contribution in the conductivity of 98% and $\mu_i N_i = 2.6 \times 10^{10}$ cm$^{-1}$V$^{-1}$s$^{-1}$ is found at moderate sub-bandgap illumination, which may slightly deviate from the values in the dark. The ion impact in conductivity decreases to 65% under extensive illumination. Although photoexcitation enhances the ionic conductivity of OMHPs by several orders of magnitude[74] the photo-induced free hole carriers still dominate in the charge transport with a much higher hole generation rate which is in agreement with our calculations. The presence of ions or mobile defects can lead to deformation of the electric field profile inside MAPbBr$_3$ and to its inevitable polarization. This is in agreement with the observed ion migration in many other studies[31,38,55,75–77] of OMHPs. It was also theoretically shown by Apiroz *et al.*[19] that mobile defect vacancies can quickly drift to the metal contact on time scales of ~1 μs, but slower ions migration was experimentally observed taking minutes or even hours.

The relatively low capture cross section found by PHES in the dynamic regime can be explained by screening from the high relative permittivity of OMHPs[18] leading to a low SRH recombination rate. The small capture cross section of defects was previously ascribed by polaronic nature of charge carriers in OMHPs and the rearrangement of MA dipoles in the vicinity of defects[18]. The defect screening produced by ions was also proposed by several groups[5,62,78].

To conclude, we characterize the electrical structure of MAPbBr$_3$ which has been hardly described so far. Three deep levels with activation energies $E_V$ + 1.05 eV, $E_V$ + 1.5 eV, and $E_V$ + 1.9 eV (or $E_C$ - 1.9 eV) were detected. We emphasize that in our study DLs are founded by steady-state PHES allowing the direct visualization of deep states by the free carriers' excitation without intermediate complicated calculations. We showed that PHES with enhanced illumination is the important tool which allows deep levels defect detection even in the material with low defect concentration inaccessible by other spectroscopy methods. Relatively low hole capture cross section of $\sigma_h = 4 \times 10^{-17}$ cm$^2$ found by photoconductivity relaxations signify SRH lifetime of $10^{-4}$-$10^{-1}$ $s$ in MAPbBr$_3$ single crystals. The value of the hole mobility 162±17 cm$^2$V$^{-1}$s$^{-1}$ and the hole lifetime 140 µs were found by ToF.

We also found that ionic (or another low mobile species) conductivity prevails in the dark and the electronic conductivity prevail under ionic part during illumination due to the higher free hole generation rate. This study opens a richer picture of the role and properties of deep levels in MAPbBr$_3$ single crystals for the first time. The deeper knowledge of the electrical structure of OMHP could open further opportunities in development of more feasible technology. The discovered deep levels and low mobile ions/defect can participate in the polarization of bulk semiconductor and influence free charge transport properties of OMHP devices. Strategies can consider to modify the growth of OMHPs by controlling the redox chemistry of halide under mild oxidation as well as halide doping, modifying precursors ratio, precursors impurity and growth temperature to further restrict the deep levels.

# Supplementary Information

## Deep Levels and Mixed Conductivity in Organometallic Halide Perovskites

Artem Musiienko[1*], Pavel Moravec[1], Roman Grill[1], Petr Praus[1], Igor Vasylchenko[1], Jakub Pekarek[1], Jeremy Tisdale[2], Katarina Ridzonova[1], Eduard Belas[1], Lucie Abelová[3,4], Bin Hu[2], Eric Lukosi[5], Mahshid Ahmadi[2*]

## Methods

### Photo Hall effect spectroscopy measurements

To study electrical properties of the single crystal MAPbBr$_3$ we used photo-Hall effect spectroscopy with enhanced monochromatic illumination. The Hall voltage can found under the steady-state conditions. The longitudinal voltage *V*, the current *I*, and the transverse Hall voltage $V_H$ were measured directly from the experiment. The conductivity $\sigma$, Hall coefficient $R_H$ and the Hall mobility $\mu_H$ were obtained by the following relations:

$$\sigma = \frac{I \cdot L}{V \cdot S} \quad (10)$$

$$R_H = \frac{V_H d}{I \cdot B}, \quad (11)$$

$$\mu_H = \sigma \cdot R_H = \frac{L}{S \cdot V} \cdot \frac{V_H d}{B}, \quad (12)$$

where *S*, *L*, and *d* are the cross section, the distance between conductivity probes, and thickness of the sample, respectively. Note that the obtained value of $\mu_H$ is determined by $V_H/V$ ratio. Therefore, $\mu_H$ remains unaffected by contingent transversal inhomogeneity of the sample induced for example by surface leakage current.

The photoconductivity is defined as the difference between the conductivity $\sigma(I)$ under the photon flux *I* and dark conductivity $\sigma(0)$

$$PhC = \sigma(I) - \sigma(0) \quad (13)$$

To generate free carriers in MAPbBr$_3$ we used a powerful white laser with a maximal filtered output photon flux of $1.3\times10^{15}$ cm$^{-2}$s$^{-1}$. The laser source provides nearly constant photon flux in the 0.5 - 2.4 eV energy region[79] which is in the strongest interest for DL spectroscopy. The sharp rise of the laser intensity near 1090 nm (spectral width of 25 nm) [79] was skipped by a monochromator. The spectral resolution of the monochromator was measured using Ocean Optics spectrometer. The resolution (defined by FWHM) grows nearly linearly from 0.01 eV at 0.5 eV up to 0.15 eV at 2.3 eV. Linear interpolation of these values was used to correctly determine the DL threshold energies at intermediate energies. Samples with typical dimensions of 1×2.5×9 mm$^3$ were used for galvanomagnetic measurements.

The PhC spectra in a linear scale are shown in **Figure S1**. As can be seen the initial photon energies started from 0.6 eV when the first increase was detected at 1.05 eV. The second threshold is detected at 1.48 eV.

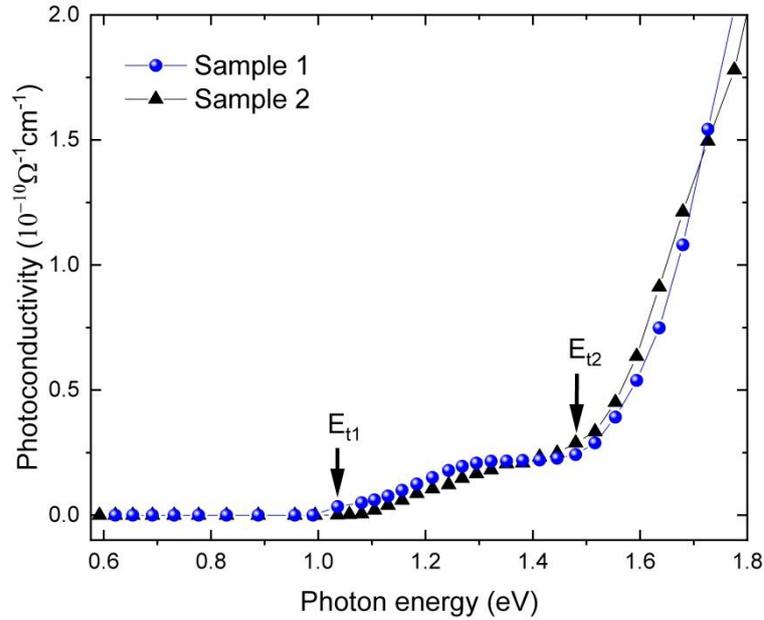

**Figure S1** Photoconductivity spectra of two single crystals of MAPbBr$_3$ as a function of photon energy. Vertical arrows show the deep level threshold energy. Spectra are plotted in linear scale to show low photon energy data where no photoconductivity response was detected.

**Hall effect in material with electrical and mixed conductivity:**

The increase of Hall coefficient and Hall signal enhancement can be explained by ions contribution as was qualitatively explained in Eq. (3) and Eq. (4). Here we have derived the ion contribution in detail:

The Hall signal in a semiconductor material without mixed conductivity follows below equation:

$$R_{H1} = \frac{V_{H1} d}{IB} = \frac{1}{q_e p} \tag{14}$$

The Hall voltage $V_{H1}$ is defined by the hole concentration $p$. If now we consider a semiconductor with mixed conductivity, Eq. (14) is modified to:

$$R_{H2} = \frac{V_{H2}d}{I \cdot B} = \frac{1}{q_e p \left(1 + \frac{\mu_i N_i}{\mu_p p}\right)^2} \quad (15)$$

As can be seen, the Hall signal is suppressed by the ionic impact $\mu_i N_i$:

$$\frac{1}{q_e p} > \frac{1}{q_e p \left(1 + \frac{\mu_i N_i}{\mu_p p}\right)^2} \quad (16)$$

The Hall signal in a semiconductor with pure electronic conductivity is higher ($V_{H1} > V_{H2}$) however, the Hall signal can be enhanced by the light illumination by increasing hole concentration as it was shown in the manuscript.

Here we show $R_{H1}$ and $R_{H2}$ as a function of hole concentration in **Figure S2**. The value of $\mu_i N_i$ product is fixed to $10^{11}$ cm$^{-1}$V$^{-1}$s$^{-1}$ for simplicity. As it can be seen, the generation of free holes lead to an increase in Hall coefficient while ion free Hall coefficient decreases.

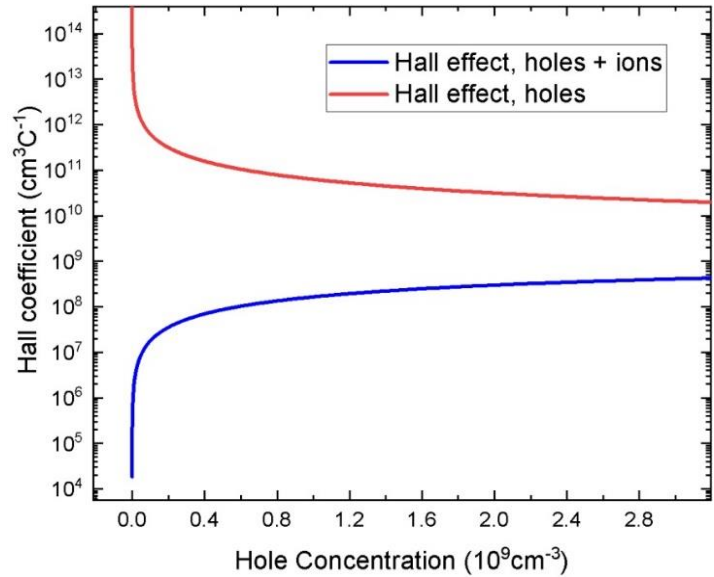

**Figure S2:** Plot of Hall coefficient as a function of holes concentration.

Consequently during photoexcitation when the hole concentration dominant the ion concentration, the Hall mobility is increased by free hole generation as seen in **Figure S2**. Note that Hall mobility does not directly represent drift mobility as shown by Eq. (4). To explicitly show this we have plotted $\mu_H$ as a function of hole concentration (according to Eq. (4)).

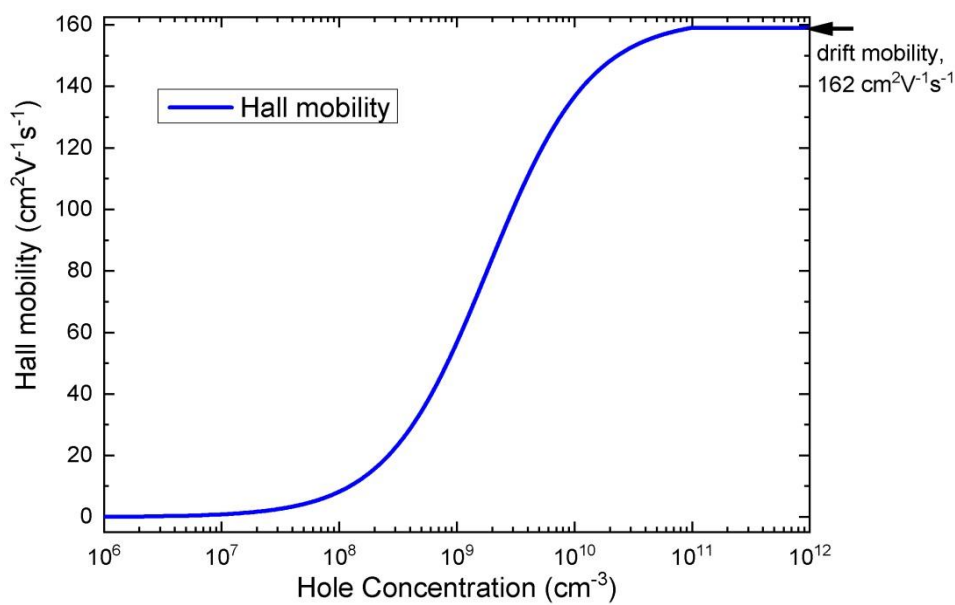

**Figure S3:** Hall mobility as a function of hole concentration in semiconductor with mixed conductivity type.

As can be seen in **Figure S3**, Hall mobility starts from very low values <1 cm$^2$V$^{-1}$s$^{-1}$ and then converges to hole drift mobility at higher hole concentration. Note that it is well established that ion concentration $Ni$ can also enhance under illumination in OMHPs[66], but as it was shown in the manuscript this change is two order of magnitude lower than the impact of free hole in photoconductivity. Therefore, a slight increase in the ion concentration does not influence PHES deep level detection.

**Materials**

Lead bromide (≥98%) and methylammonium bromide (MABr) were purchased from Alfa Aesar, Dimethylformamide (DMF) (anhydrous, 99.8%) were purchased from Sigma Aldrich. All precursors and solvents were used as received. Single crystals of MAPbBr$_3$ was grown via an inverse temperature crystallization (ITC) method[80] with a slightly modified temperature gradient. In this method, MABr and PbBr$_2$ were mixed with 1:0.8 molar ratio in 1 mL DMF solvent. The solution was stirred for 1 h and then filtered using a 0.2 μm PTFE filter. The solution was then placed in an oil bath on a hot plate and MAPbBr$_3$ single crystals are grown by slow temperature increase during overnight to 75 °C. After growth, the crystals were washed using dichloromethane (DCM). From UV-Visible absorption spectrum a sharp edge is observed at 540 nm (**Figure S4**) with the extrapolated bandgap energy of around 2.2 eV, in agreement with literature[80]. The primary photoluminescence emission peak is centered at 530 nm, and a secondary peak is centered at 575 nm due to strong self-absorption, which are consistent with published results[81]. Prior to contacts deposition, MAPbBr$_3$ samples were initially rinsed in pure toluene and then dried in using compressed dry air. The MAPbBr$_3$ single crystals were prepared in two different ways. The first sample was glued on a silicone support with a commercially available acrylate lacquer (30 wt. % in toluene). The front crystal face was mechanically covered by a mask made of an aluminum foil. The second sample was masked using only acrylate lacquer. The latter was dried by keeping crystals in the open air for 10 minutes. Afterward, gold contacts were evaporated in 10$^{-5}$ mbar vacuum; the resulting samples were taken off the supports in pure toluene, additionally rinsed in the same solution and dried in air flow. Between measurements to prevent material degradation we kept crystals in dark condition in sealed containers over both moisture and oxygen absorbents. Samples typical resistivity of 3x10$^8$ Ωcm was found from four point resistivity measurements.

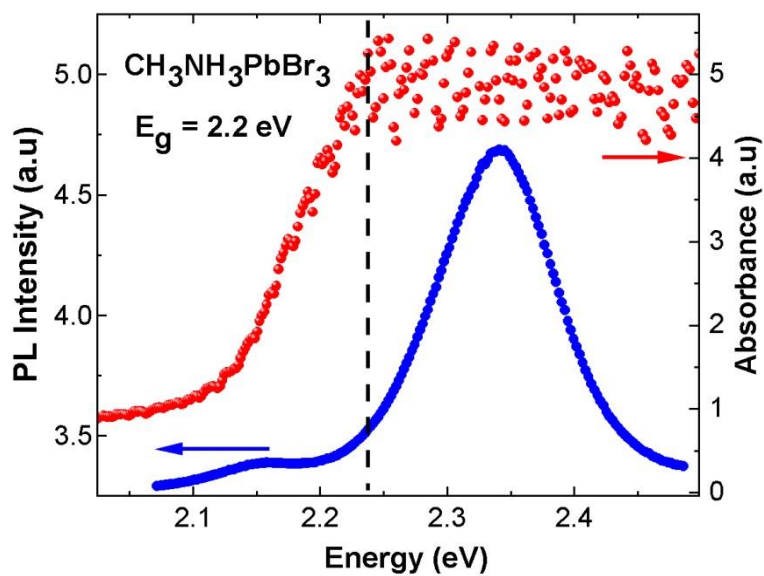

**Figure S4.** UV-Visible absorption (red circles) and PL spectra (blue circles) of MAPbBr$_3$ single crystals.

In the second method MAPbBr$_3$ single crystals (sample 3) were grown by dissolving PbBr$_2$ and MABr in 2.5 ml DMF at 55 ° C for 1-1.5 h. Then, 75 μl of HCOOH was added to the solution, and after dissolution of the formed precipitate, the solution was filtered through a syringe filter with a pore size of 0.2 μm. The solution (1) was heated to 70 ° C for 15 minutes and then to 73 ° C for 40 minutes. During this time, many small crystals formed in solution as seed crystals. Next the seed crystals were placed in fresh preheated solution (2), where they were allowed to grow larger at 55 ° C for 19 h. This cycle was repeated 2 more times for growing larger crystals. The grown crystals were then dried with Kimwipe. The whole crystallization was performed in the glovebox. Samples typical resistivity of $5 \times 10^7$ Ωcm was found from four point resistivity measurements.

**Photo-Hall in MAPbBr$_3$ single crystals grown with different method:**

Deep levels with activation energies $E_V + 1.05$ eV, $E_V + 1.5$ eV, and $E_V + 1.9$ eV ($E_C - 1.9$ eV) were detected by PHES in MAPbBr$_3$ as shown in **Figure S5.** This sample was grown with slightly different method in the lab of Czech Technical University in Prague (Details can be found in materials section).

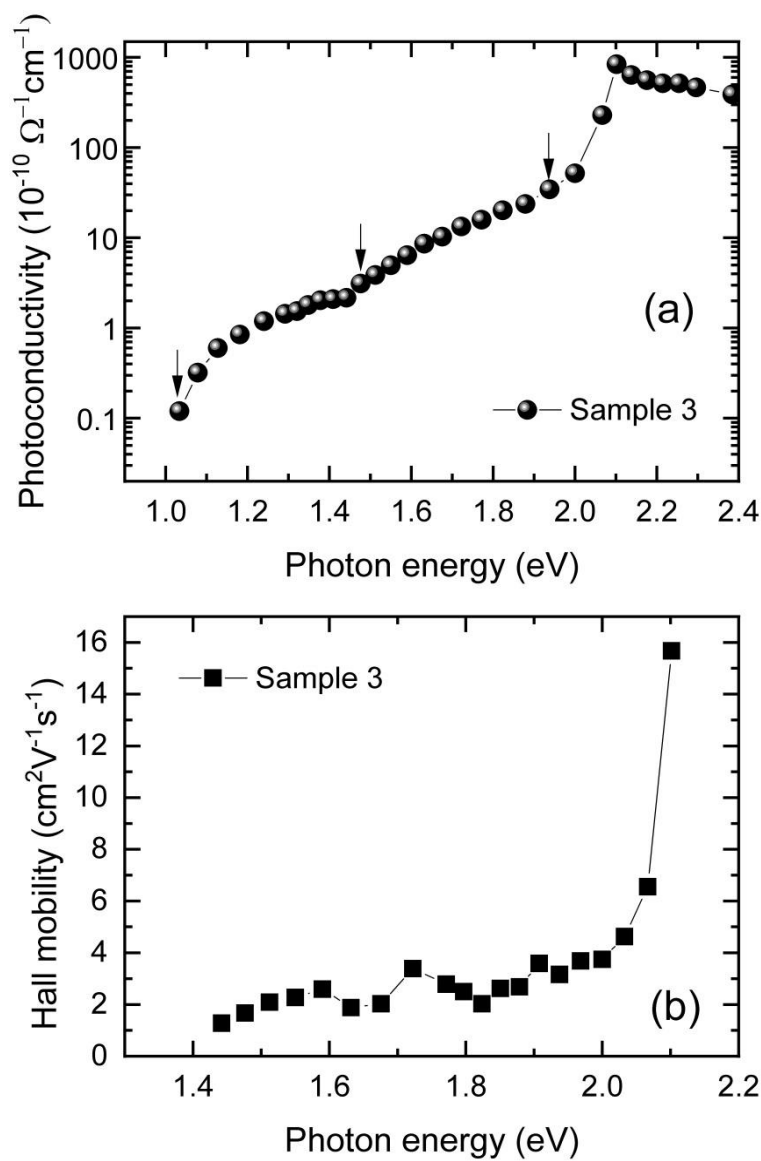

**Figure S5:** (a) Photoconductivity and (b) Hall mobility spectra of MAPbBr$_3$ single crystals grown in another laboratory as a function of photon energy. Vertical arrows show the deep level threshold energy.

**Time of flight measurements**

MAPbBr$_3$ single crystals with a typical thickness of 1.8 mm are studied by laser-induced transient current measurements. For this experiment, two opposite gold contacts were deposited on single crystal MAPbBr$_3$ by thermal evaporation. A DC bias, varying between 6 V and 12 V, was applied to the sample. Anode side of the sample was illuminated by the above-bandgap pulse laser, that was powered by the amplified output of the arbitrary waveform generator. Only hole signal was collected. Incident probe pulses had the wavelength of 450 nm, pulse width of 100 ns, and the repetition rate was set to 100 Hz. We applied neutral density optical filter to attenuate the light pulse peak power to ~ 5 mW. So that, the used probe pulse intensity is weak enough not to affect the properties of the studied material. We use an in-house designed wide bandwidth high frequency voltage amplifier to record CWF by the digital sampling oscilloscope (4 GHz, 11 bit resolution). The detailed scheme of the ToF apparatus was shown in our previous works[25,82].

**Shockley-Read-Hall recombination theory**

The Shockley-Read-Hall hole recombination-generation rate[83] complemented by illumination-mediated deep level - band transitions is given by relations

$$\frac{\partial p}{\partial t} = -\sum_i U_i^h + \sum_i I_{vi}, \tag{17}$$

$$U_i^h = \sigma_{hi} v_h [n_{ti} p - (N_{ti} - n_{ti}) p_{1i}], \tag{18}$$

$$I_{vi} = I \tilde{\alpha}_{hi} (N_{ti} - n_{ti}). \tag{19}$$

where $n$, $p$, $n_{ti}(p_{ti},)$, $U_i^h$, are the densities of free electrons, free holes, electrons(holes) trapped in the $i$-th level, and hole net recombination rate at the $i$-th level. The quantities defining recombination rates $n_{ti}$, $N_{ti}$, $\sigma_{hi}$, and $v_h$ in Eqs. (18-19) are intrinsic carrier density, $i$-th DL density, hole thermal capture cross section, and hole thermal velocity. Symbol $p_{1i}$ stands for electron and hole densities in case of Fermi level $E_F$ being set equal to the DL ionization energy $E_{ti}$[83]. The effect of illumination on the $i$-th DL occupancy is defined by $I_{ci}$ ($I_{vi}$) generation rate from $i$-th level to the conduction (valence) band where $I$, $\tilde{\alpha}_{ei}$, and $\tilde{\alpha}_{hi}$ are the photon flux and photon capture cross sections relevant to the conduction and valence band transition respectively. The hole lifetimes $\tau$ can be found by relation

$$\tau = \frac{1}{\sum_i v_h \sigma_{hi} n_{ti}} \tag{20}$$

The inter-band light induced generation rate and bimolecular recombination rate are neglected in the case of sub-bandgap illumination. The simplified version of Eqs. (18-19) in the case of single DL is used to find the hole capture cross section in this paper.

**Authors Contribution:**

A.M. and M.A. conceived the project and A.M. performed experiments directed by M.A. J.T. prepared samples in the lab in US. and I.V. prepared metal contacts and controlled sample and contact quality for electrical measurements. A.M. performed photo-Hall effect measurements (PHES), data analysis and theoretical calculations with assistance from P.M. and R.G., R.G. contributed in deep data analysis and theoretical models based on Shockley Reed Hall differential equations. P. P. performed Time of flight (ToF) measurements with help from J.P. and E.B., P.P. and R.G. assisted A.M. in ToF data analysis. K.R. contributed in ToF and PHES measurements for long timescales. L.A. prepared sample in Czeck Republic with slightly different method from J.T. to ensure data universalness. M.A. performed UV-Vis and PL spectroscopy. A.M. and M.A. wrote the manuscript. All co-authors contributed to experimental design and data analysis and revised the manuscript critically for important intellectual content and approved the final version.

**Acknowledgment**

J. T., B. H., E. L. and M. A., acknowledge financial support from US Department of Homeland Security (grant # 2016-DN-077-ARI01). A. M., P. M., I. V., J. P., K. R. and E. B. acknowledge financial support from the Grant Agency of Charles University, projects No. 1374218 and No. 1230218 and Grant agency of the Czech Republic, Grant no. GJ17-26041Y. Part of this work was conducted in the Micro-Processing Research Facility, a University of Tennessee Core Facility.

**Disclaimer:** The views and conclusions contained in this document are those of the authors and should not be interpreted as necessarily representing the official policies, either expressed or implied, of the U.S. Department of Homeland Security.

**References**

[1] A. Kojima, K. Teshima, Y. Shiral, M. Tsutomu, *J. Am. Chem. Soc.* **2009**, *131*, 6050.
[2] N. G. Park, *Mater. Today* **2015**, *18*, 65.
[3] L. Zhang, X. Yang, Q. Jiang, P. Wang, Z. Yin, X. Zhang, H. Tan, Y. (Michael) Yang, M. Wei, B. R. Sutherland, E. H. Sargent, J. You, *Nat. Commun.* **2017**, *8*, 15640.
[4] M. Ahmadi, T. Wu, B. Hu, *Adv. Mater.* **2017**, *29*, 1605242.
[5] H.-H. Fang, S. Adjokatse, H. Wei, J. Yang, G. R. Blake, J. Huang, J. Even, M. A. Loi, *Sci. Adv.* **2016**, *2*, e1600534.
[6] X. Liu, H. Zhang, B. Zhang, J. Dong, W. Jie, Y. Xu, *J. Phys. Chem. C* **2018**, *122*, 14355.
[7] K. Miyano, N. Tripathi, M. Yanagida, Y. Shirai, *Acc. Chem. Res.* **2016**, *49*, 303.
[8] B. Sopori, *J. Electron. Mater.* **2005**, *31*, 972.
[9] W. J. Yin, T. Shi, Y. Yan, *Appl. Phys. Lett.* **2014**, *104*, 063903.
[10] A. Walsh, D. O. Scanlon, S. Chen, X. G. Gong, S. Wei, *Angew. Chemie* **2014**, *54*, 1791.
[11] A. Walsh, A. Zunger, *Nat. Mater.* **2017**, *16*, 964.
[12] J. Kim, S. H. Lee, J. H. Lee, K. H. Hong, *J. Phys. Chem. Lett.* **2014**, *5*, 1312.

[13] I. L. Braly, D. W. DeQuilettes, L. M. Pazos-Outón, S. Burke, M. E. Ziffer, D. S. Ginger, H. W. Hillhouse, *Nat. Photonics* **2018**, *12*, 355.
[14] O. von Roos, P. T. Landsberg, *J. Appl. Phys.* **1985**, *57*, 4746.
[15] A. Buin, R. Comin, J. Xu, A. H. Ip, E. H. Sargent, *Chem. Mater.* **2015**, *27*, 4405.
[16] A. Buin, P. Pietsch, J. Xu, O. Voznyy, A. H. Ip, R. Comin, E. H. Sargent, *Nano Lett.* **2014**, *14*, 6281.
[17] M. L. Agiorgousis, Y. Y. Sun, H. Zeng, S. Zhang, *J. Am. Chem. Soc.* **2014**, *136*, 14570.
[18] I. Anusca, S. Balčiūnas, P. Gemeiner, Š. Svirskas, M. Sanlialp, G. Lackner, C. Fettkenhauer, J. Belovickis, V. Samulionis, M. Ivanov, B. Dkhil, J. Banys, V. V. Shvartsman, D. C. Lupascu, *Adv. Energy Mater.* **2017**, *7*, 1700600.
[19] J. M. Azpiroz, E. Mosconi, J. Bisquert, F. De Angelis, *Energy Environ. Sci.* **2015**, *8*, 2118.
[20] A. Castaldini, A. Cavallini, B. Fraboni, P. Fernandez, J. Piqueras, *J. Appl. Phys.* **1998**, *83*, 2121.
[21] J. Zazvorka, P. Hlidek, R. Grill, J. Franc, E. Belas, *J. Lumin.* **2016**, *177*, 71.
[22] J. Franc, P. Hlídek, E. Belas, V. Linhart, S. Pospíšil, R. Grill, *IEEE Trans. Nucl. Sci.* **2005**, *52*, 1956.
[23] A. Carvalho, A. K. Tagantsev, S. Öberg, P. R. Briddon, N. Setter, *Phys. Rev. B* **2010**, *81*, 075215.
[24] J. Pousset, I. Farella, S. Gambino, A. Cola, *J. Appl. Phys.* **2016**, *119*, 105701.
[25] A. Musiienko, R. Grill, J. Pekárek, E. Belas, P. Praus, J. Pipek, V. Dědič, H. Elhadidy, *Appl. Phys. Lett.* **2017**, *111*, 082103.
[26] B. Wenger, P. K. Nayak, X. Wen, S. V. Kesava, N. K. Noel, H. J. Snaith, *Nat. Commun.* **2017**, *8*, 590.
[27] Y. Chen, H. T. Yi, X. Wu, R. Haroldson, Y. N. Gartstein, Y. I. Rodionov, K. S. Tikhonov, A. Zakhidov, X. Y. Zhu, V. Podzorov, *Nat. Commun.* **2016**, *7*, 12253.
[28] Y. Chen, H. T. Yi, V. Podzorov, *Phys. Rev. Appl.* **2016**, *5*, 034008.
[29] A. Musiienko, R. Grill, P. Hlídek, P. Moravec, E. Belas, J. Zázvorka, G. Korcsmáros, J. Franc, I. Vasylchenko, *Semicond. Sci. Technol.* **2016**, *32*, 015002.
[30] A. Musiienko, R. Grill, P. Moravec, G. Korcsmáros, M. Rejhon, J. Pekárek, H. Elhadidy, L. Šedivý, I. Vasylchenko, *J. Appl. Phys.* **2018**, *123*, 161502.
[31] Y. Yang, M. Yang, Z. Li, R. Crisp, K. Zhu, M. C. Beard, *J. Phys. Chem. Lett.* **2015**, *6*, 4688.
[32] S. Shrestha, G. J. Matt, A. Osvet, D. Niesner, R. Hock, C. J. Brabec, *J. Phys. Chem. C* **2018**, *122*, 5935.
[33] I. Supplementary, *Nat. Photonics* **2015**, *9*, 444.
[34] Y. C. Kim, K. H. Kim, D. Y. Son, D. N. Jeong, J. Y. Seo, Y. S. Choi, I. T. Han, S. Y. Lee, N. G. Park, *Nature* **2017**, *550*, 87.
[35] K. Seeger, *Semiconductor Physics*, Springer-Verlag/Wien, Wien, **1973**.
[36] H. Wei, D. DeSantis, W. Wei, Y. Deng, D. Guo, T. J. Savenije, L. Cao, J. Huang, *Nat. Mater.* **2017**, *16*, 826.
[37] L. Cojocaru, S. Uchida, K. Tamaki, P. V. V. Jayaweera, S. Kaneko, J. Nakazaki, T. Kubo, H. Segawa, *Sci. Rep.* **2017**, *7*, 11790.
[38] N. Ahn, K. Kwak, M. S. Jang, H. Yoon, B. Y. Lee, J. K. Lee, P. V. Pikhitsa, J. Byun, M. Choi, *Nat. Commun.* **2016**, *7*, 13422.
[39] M. S. Dresselhaus, *SOLID STATE PHYSICS PART I Transport Properties of Solids*, **2001**.
[40] J. Wiktor, F. Ambrosio, A. Pasquarello, *J. Mater. Chem. A* **2018**, *6*, 16863.
[41] W. A. (Walter A. Harrison, *Electronic Structure and the Properties of Solids : The Physics of the Chemical Bond*, Dover Publications, **1989**.
[42] A. Zerrai, G. Marrakchi, G. Bremond, *J. Appl. Phys.* **2000**, *87*, 4293.
[43] J. Blakemore, *Phys. Rev.* **1967**, *163*, 809.

[44] D. Guo, D. Bartesaghi, H. Wei, E. M. Hutter, J. Huang, T. J. Savenije, *J. Phys. Chem. Lett.* **2017**, *8*, 4258.

[45] J. W. Rosenberg, M. J. Legodi, Y. Rakita, D. Cahen, M. Diale, *J. Appl. Phys.* **2017**, *122*, 145701.

[46] A. Musiienko, R. Grill, J. Pekárek, E. Belas, P. Praus, J. Pipek, V. Dědič, H. Elhadidy, *Appl. Phys. Lett.* **2017**, *111*, 082103.

[47] P. Praus, E. Belas, J. Franc, R. Grill, P. Hoschl, J. Pekarek, *IEEE Trans. Nucl. Sci.* **2014**, *61*, 2333.

[48] A. D. Wright, C. Verdi, R. L. Milot, G. E. Eperon, M. A. Pérez-Osorio, H. J. Snaith, F. Giustino, M. B. Johnston, L. M. Herz, *Nat. Commun.* **2016**, *7*, 11755.

[49] K. Miyata, T. L. Atallah, X.-Y. Zhu, *Sci. Adv.* **2017**, *3*, e1701469.

[50] Y.-B. Lu, X. Kong, X. Chen, D. G. Cooke, H. Guo, *Sci. Rep.* **2017**, *7*, 41860.

[51] N. Liu, C. Yam, *Phys. Chem. Chem. Phys.* **2018**, *20*, 6800.

[52] D. Chattopadhyay, H. J. Queisser, *Rev. Mod. Phys.* **1981**, *53*, 745.

[53] R. Mansfield, *Proc. Phys. Soc. Sect. B* **1956**, *69*, 76.

[54] W. Peng, C. Aranda, O. M. Bakr, G. Garcia-Belmonte, J. Bisquert, A. Guerrero, *ACS Energy Lett.* **2018**, *3*, 1477.

[55] J. M. Frost, A. Walsh, *Acc. Chem. Res.* **2016**, *49*, 528.

[56] R. Grill, J. Franc, H. Elhadidy, E. Belas, Š. Uxa, M. Bugár, P. Moravec, P. Höschl, *IEEE Trans. Nucl. Sci.* **2012**, *59*, 2383.

[57] X. Y. Zhu, V. Podzorov, *J. Phys. Chem. Lett.* **2015**, *6*, 4758.

[58] K. Miyata, T. L. Atallah, X. Y. Zhu, *Sci. Adv.* **2017**, *3*, e1701469.

[59] V. Dědič, J. Franc, H. Elhadidy, R. Grill, E. Belas, P. Moravec, J. Zázvorka, P. Höschl, *J. Instrum.* **2013**, *8*, C01008.

[60] G. Landi, H. C. Neitzert, C. Barone, C. Mauro, F. Lang, S. Albrecht, B. Rech, S. Pagano, *Adv. Sci.* **2017**, *4*, 1700183.

[61] T. Zhao, W. Shi, J. Xi, D. Wang, Z. Shuai, *Sci. Rep.* **2016**, *7*, 19968.

[62] J. Ma, L.-W. Wang, *Nano Lett.* **2017**, *17*, 3646.

[63] S. Reynolds, M. Brinza, M. L. Benkhedir, G. J. Adriaenssens, in *Springer Handb. Electron. Photonic Mater.*, Springer International Publishing, Cham, **2017**, pp. 1–1.

[64] R. Newman, W. W. Tyler, *Solid State Phys.* **1959**, *8*, 49.

[65] G. Maculan, A. D. Sheikh, A. L. Abdelhady, M. I. Saidaminov, M. A. Haque, B. Murali, E. Alarousu, O. F. Mohammed, T. Wu, O. M. Bakr, *J. Phys. Chem. Lett.* **2015**, *6*, 3781.

[66] Y. Bi, E. M. Hutter, Y. Fang, Q. Dong, J. Huang, T. J. Savenije, *J. Phys. Chem. Lett.* **2016**, *7*, 923.

[67] Hojun Yoon, K. M. Edmondson, G. S. Kinsey, R. R. King, P. Hebert, R. K. Ahrenkiel, B. T. Cavicchi, N. H. Karam, in *Conf. Rec. Thirty-First IEEE Photovolt. Spec. Conf. 2005.*, IEEE, **n.d.**, pp. 842–845.

[68] R. H. Bube, *J. Appl. Phys.* **1993**, *74*, 5138.

[69] F. Stockmann, *Phys. Status Solidi* **1969**, *34*, 751.

[70] A. Musiienko, R. Grill, P. Moravec, P. Fochuk, I. Vasylchenko, H. Elhadidy, L. Šedivý, *Phys. Rev. Appl.* **2018**, *10*, 014019.

[71] M.-H. Du, *J. Phys. Chem. Lett.* **2015**, *6*, 1461.

[72] D. Meggiolaro, S. G. Motti, E. Mosconi, A. J. Barker, J. Ball, C. Andrea Riccardo Perini, F. Deschler, A. Petrozza, F. De Angelis, *Energy Environ. Sci.* **2018**, *11*, 702.

[73] B. Kang, K. Biswas, *Phys. Chem. Chem. Phys.* **2017**, *19*, 27184.

[74] G. Y. Kim, A. Senocrate, T.-Y. Yang, G. Gregori, M. Grätzel, J. Maier, *Nat. Mater.* **2018**, *17*, 445.

[75] S. Meloni, T. Moehl, W. Tress, M. Franckeviius, M. Saliba, Y. H. Lee, P. Gao, M. K.

Nazeeruddin, S. M. Zakeeruddin, U. Rothlisberger, M. Graetzel, *Nat. Commun.* **2016**, *7*, 10334.

[76] Y.-C. Zhao, W.-K. Zhou, X. Zhou, K.-H. Liu, D.-P. Yu, Q. Zhao, *Light Sci. Appl.* **2016**, *6*, e16243.

[77] A. J. Barker, A. Sadhanala, F. Deschler, M. Gandini, S. P. Senanayak, P. M. Pearce, E. Mosconi, A. J. Pearson, Y. Wu, A. R. Srimath Kandada, T. Leijtens, F. De Angelis, S. E. Dutton, A. Petrozza, R. H. Friend, *ACS Energy Lett.* **2017**, *2*, 1416.

[78] M. Šimėnas, J. Banys, E. E. Tornau, *J. Mater. Chem. C* **2018**, *6*, 1487.

[79] "COMPACT SUPERCONTINUUM LASER," can be found under http://www.nktphotonics.com/lasers-fibers/en/product/superk-compact-supercontinuum-lasers/, **2017**.

[80] M. I. Saidaminov, A. L. Abdelhady, B. Murali, E. Alarousu, V. M. Burlakov, W. Peng, I. Dursun, L. Wang, Y. He, G. MacUlan, A. Goriely, T. Wu, O. F. Mohammed, O. M. Bakr, *Nat. Commun.* **2015**, *6*, 7586.

[81] Y. Fang, H. Wei, Q. Dong, J. Huang, *Nat. Commun.* **2017**, *8*, 14417.

[82] P. Praus, E. Belas, J. Bok, R. Grill, J. Pekarek, *IEEE Trans. Nucl. Sci.* **2016**, *63*, 246.

[83] M. Borgwardt, P. Sippel, R. Eichberger, M. P. Semtsiv, W. T. Masselink, K. Schwarzburg, *J. Appl. Phys.* **2015**, *117*, 215702.